\documentclass{PoS}

\usepackage{epsfig}
\usepackage{graphicx}
\usepackage{txfonts}
\usepackage{multicol}
 
\def\inte{{\em INTEGRAL}} 

\title{Accreting millisecond X-ray pulsars: 10 years of INTEGRAL observations}

\ShortTitle{Accreting millisecond X-ray pulsars}

\author{\speaker{M. Falanga}$^{1}$, L. Kuiper$^{2}$, J. Poutanen$^{3}$, D. K. Galloway$^{4}$, E. Bozzo$^{5}$, A. Goldwurm$^{6,7}$, W. Hermsen$^{2}$, L. Stella$^{8}$\\ 
       $^{1}$ International Space Science Institute (ISSI), Hallerstrasse 6, CH-3012 Bern, Switzerland\\
       $^{2}$  SRON--Netherlands Institute for Space Research, Sorbonnelaan 2, 3584 CA, Utrecht, The Netherlands \\
       $^{3}$ Astronomy Division,  Department of Physics, P.O. Box 3000, FI-90014 University of Oulu, Finland\\
       $^{4}$ Monash Center for Astrophysics, School of Physics, and School of Mathematical Sciences, Monash University, VIC 3800, Australia\\
       $^{5}$ ISDC - Universit\`e de Gen\`eve. Chemin d'Ecogia 16, CH-1290 Versoix, Switzerland \\
       $^{6}$ Service d`Astrophysique (SAp), IRFU/DSM/CEA-Saclay, 91191 Gif-sur-Yvette Cedex, France\\ 
       $^{7}$ Unit\'e mixte de recherche Astroparticule et Cosmologie, 10 rue Alice Domon et Leonie Duquet, F-75205 Paris, France\\ 
       $^{8}$  IINAF--Osservatorio Astronomico di Roma, via Frascati 33, 00040 Monteporzio Catone (Roma), Italy\\
       E-mail: \email{mfalanga@issibern.ch}}

\abstract{During the last 10 years, \inte\ made a unique contribution to the study of accreting millisecond X-ray pulsars (AMXPs), discovering three of the 14 sources now known of this class. Besides increasing the number of known AMXPs, \inte\ also carried out observations of these objects above 20 keV, substantially advancing our understanding of their behaviour. We present here a review of all the AMXPs observed with \inte\ and discuss the physical interpretation of their behaviour in the X-ray domain. We focus in particular on the lightcurve profile during outburst, as well as the timing, spectral, and thermonuclear type-I X-ray bursts properties.}

\FullConference{"An INTEGRAL view of the high-energy sky (the first 10 years)"
9th INTEGRAL Workshop and celebration of the 10th anniversary of the launch,\\
		October 15-19, 2012\\
		Bibliotheque Nationale de France, Paris, France}

\begin{document}

\section{Introduction}

The INTErnational Gamma-Ray Astrophysics Laboratory ({\it INTEGRAL}) was launched on 2002 October 17 to observe extreme astrophysical events in the gamma-ray sky. 
During the first ten years in orbit, {\it INTEGRAL} contributed to the discovery of three accreting millisecond X-ray pulsars (AMXPs):  
IGR~J00291+5934 in 2004 \cite{shaw05,mfb05}, IGR~J17511--3057 in 2009 \cite{mfalanga11} , and IGR~J17498--2921 in 2011 \cite{mfalanga12}. 
\inte\ also contributed to the understanding of two recently discovered AMXPs: HETE~1900.1-2455 in 2005 \cite{mfc07} and Swift~J1749.4--2807 in 2010 
\cite{ferrigno11}. XTE J1807--294 was the first AMXP serendipitously observed with \inte\ IBIS/ISGRI between February and March 2003 \cite{mfa05}. 
The AMXP SAX~J1808.4--3658 was also observed serendipitously from the beginning of its outburst in 2008 until back to quiescence 
(i.e., between September 18 and October 19, 2008). These \inte\ data still remain unpublished. The first outburst of the AMXP XTE~J1751--305 was observed 
in April 2002 and lasted 12 days \cite{Markwardt02}. \inte\ observed the second outburst from this source in March 2005 \cite{Grebenev05}, 
the third in April 2007 \cite{falangaatel07}, and the fourth in October 2009 \cite{Chenevez09,mfalanga11}. The outbursts from XTE~J1751-305 
were all relatively short and similar to each other (see \cite{Riggio11} and reference therein). 

At present, 14 AMXPs are known. All these sources are X--ray transients, i.e. they spend most of the time in a quiescent phase (X--ray luminosities 
of $10^{31}-10^{33}$ erg s$^{-1}$) and sometimes undergo X--ray outbursts reaching luminosities of $10^{36}-10^{37}$ erg s$^{-1}$. During these events 
coherent pulsations are observed with frequencies in the range between 182 and 599 Hz (see e.g. \cite{w06,p06,Patruno2012} for a review). Since the launch in 
2002, \inte\ provided a contribution to our understanding of the AMXPs by performing at least one observation of these objects 
per year. In this contribution we review all the \inte\ results on the AMXPs achieved after ten years in orbit.  
 
\section{The {\em INTEGRAL} imaging of AMXPs}

Thanks to the unprecedented {\em INTEGRAL} capabilities to image the hard X-ray sky at relatively high angular resolutions 
($12'$ for ISGRI and $3'$ for JEM-X), all AMXPs could be clearly identified in the mosaicked images at discovery and their X-ray emission unambiguously 
disentangled from that of other nearby hard X-ray emitters \cite{mfa05,mfb05,mfc07,mfalanga11,ferrigno11,mfalanga12}. An example of mosaicked sky image around an AMXP in outburst is shown in Fig. 1 ({\it left panel}).

These capabilities were particularly relevant for the observation of the source IGR~J00291+5934, which is located $\sim18'$ from the close-by persistent intermediate polar V609~Cas \cite{mfb05}. 
The AMXP source XTE~J1751--305 underwent a short period of enhanced X-ray activity during the decay of the outburst from the AMXP IGR~J17511--3057 
located $\sim20'$ away from the former object \cite{mfalanga11}. 
In the case of the AMXP IGR~J17498--2921 \inte\ was able to disentangle the contributions of the surrounding objects and to clearly distinguish between 
the type-I X-ray bursts emitted by the source and those going off in the near-by bursters SLX~1744--300/299 and 1A~1742--294 \cite{mfalanga12}.  These sources are located at $0.9^{\circ}$ and $0.86^{\circ}$ from the IGR~J17498--2921 position (see Fig. 1; {\it left panel}). 
\begin{figure}
\centering 
\includegraphics[width=0.45\textwidth]{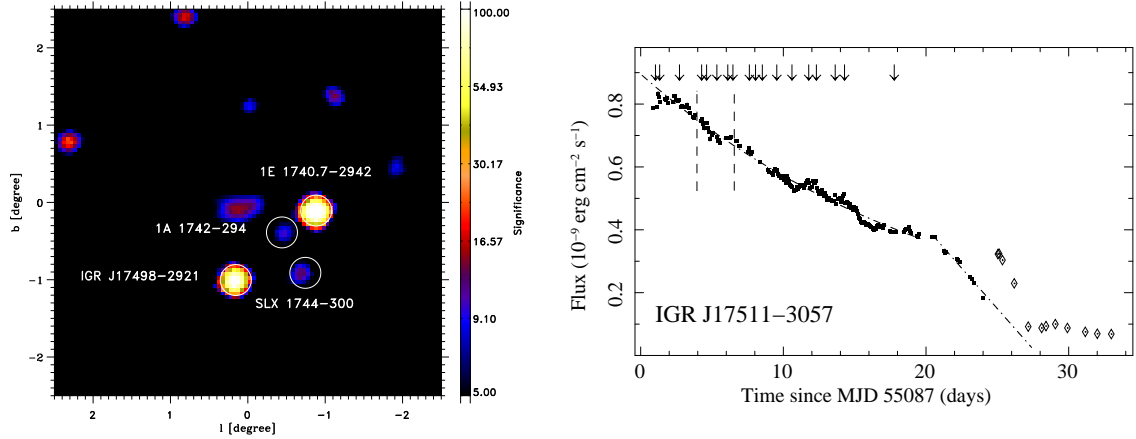}
\hspace{0.55cm}
\includegraphics[angle=-90,width=0.50\textwidth]{fig2.ps}
\vspace{0.25cm}
\caption{  \footnotesize {\it Left:} {\it INTEGRAL}/ISGRI sky image of the field of view around IGR~J17498--2921
 in the 20--100 keV band for an effective exposure of 210 ks (see \cite{mfalanga11}).  {\it Right:} {\em RXTE}/PCA (2--20 keV) outburst 
 light curve of IGR~J17511--3057. The diamonds correspond to observations 
in which both XTE~J1751--305 and IGR~J17511-3057 were active. The arrows indicate the times of the detected
X-ray bursts. The vertical dashed lines indicate the interval of the \inte\ observations (see \cite{mfalanga12}).}
\label{fig:fig1}
\end{figure} 

\section{The AMXP outburst lightcurves}

Most of the AMXPs displayed in the past very similar behaviours while in outburst. A representative lightcurve of an outburst from an AMXP is shown in Fig. 1 ({\it right panel}). The lightcurves of these events are usually 
characterized by a fast rise (a few days at the most) and an exponential decay. The latter terminates with a break, 
after which a linear decay of the source X-ray flux is observed down to the quiescence level (see e.g., \cite{mfa05,mfb05,mfc07,mfalanga11,ferrigno11,mfalanga12}). 
These outburst profiles are commonly interpreted in terms of the disk instability model, taking into account the irradiation of the disk by the central X-ray source \cite{King98}. This model has been applied to a sample of different LMXBs, 
including six AMXPs \cite{Powell07,mfalanga11,ferrigno11,mfalanga12}. In these works it was shown that the timescale of the decay light curve and its luminosity at a characteristic time are linked to the outer radius of the accretion
disk. The break (``knee'') observed during the X-ray flux decay at the end of the outburst is believed to be a consequence of mass transfer onto the outer edge of the disk, as this supply of material to the compact object is effectively cut-off when the outer disk enters the cool low-viscosity state. The knee occurs at the lowest value of the X-ray luminosity at which the outer disk edge 
can still be kept hot by central illuminating source. The measured value of this luminosity can be used to estimate for the outer disk radius. A second independent estimate of this radius is given by the timescale of the exponential decay of the lightcurve. 

\section{The AMXP X-ray spectra}

The available {\it INTEGRAL} JEM-X+ISGRI broad-band X-ray spectra of all AMXPs (XTE J1807--294, IGR~J00291+5934, HETE~1900.1--2455, IGR~J17511--3057,  Swift~J1749.4--2807, and IGR~J17498--2921), can be usually well fit by using a thermal Comptonization model {\sc compps} \cite{ps96} 
in the slab geometry \cite{mfa05,mfb05,mfc07,mfalanga11,ferrigno11,mfalanga12}. The main model parameters are the Thomson optical depth 
$\tau_{\rm T}\sim 1-2$ across the slab, the electron temperature $kT_{\rm e}\sim 25-50$ keV,
the temperature $kT_{\rm bb}\sim 0.5-1.0$ keV of the soft-seed blackbody photons assumed to be
injected from the bottom of the slab, the emission area $A_{\rm bb}\sim 20$ km$^2$. 
In Fig. 2 ({\it left panel}) we show a representative AMXP unfolded spectrum and the corresponding best-fit {\sc compps} model. 

The spectra of AMXPs are very similar to those of the so-called ``atoll'' sources at low luminosities, where the X-rays are probably produced in the boundary/spreading 
layer near the NS equator (see e.g., \cite{ps96}). These similarities can be explained by assuming that in both classes of sources the energy dissipation occurs in an optically thin medium (i.e. the accretion shock and boundary/spreading layer) and the spectral properties are determined solely by energy balance and feedback from 
the NS surface (which in turns provides cooling in the form of soft photons).  
\begin{figure}
\centering 
\includegraphics[angle=-90,width=0.45\textwidth]{fig3.ps}
\hspace{0.5cm}
\includegraphics[angle=90,width=0.50\textwidth]{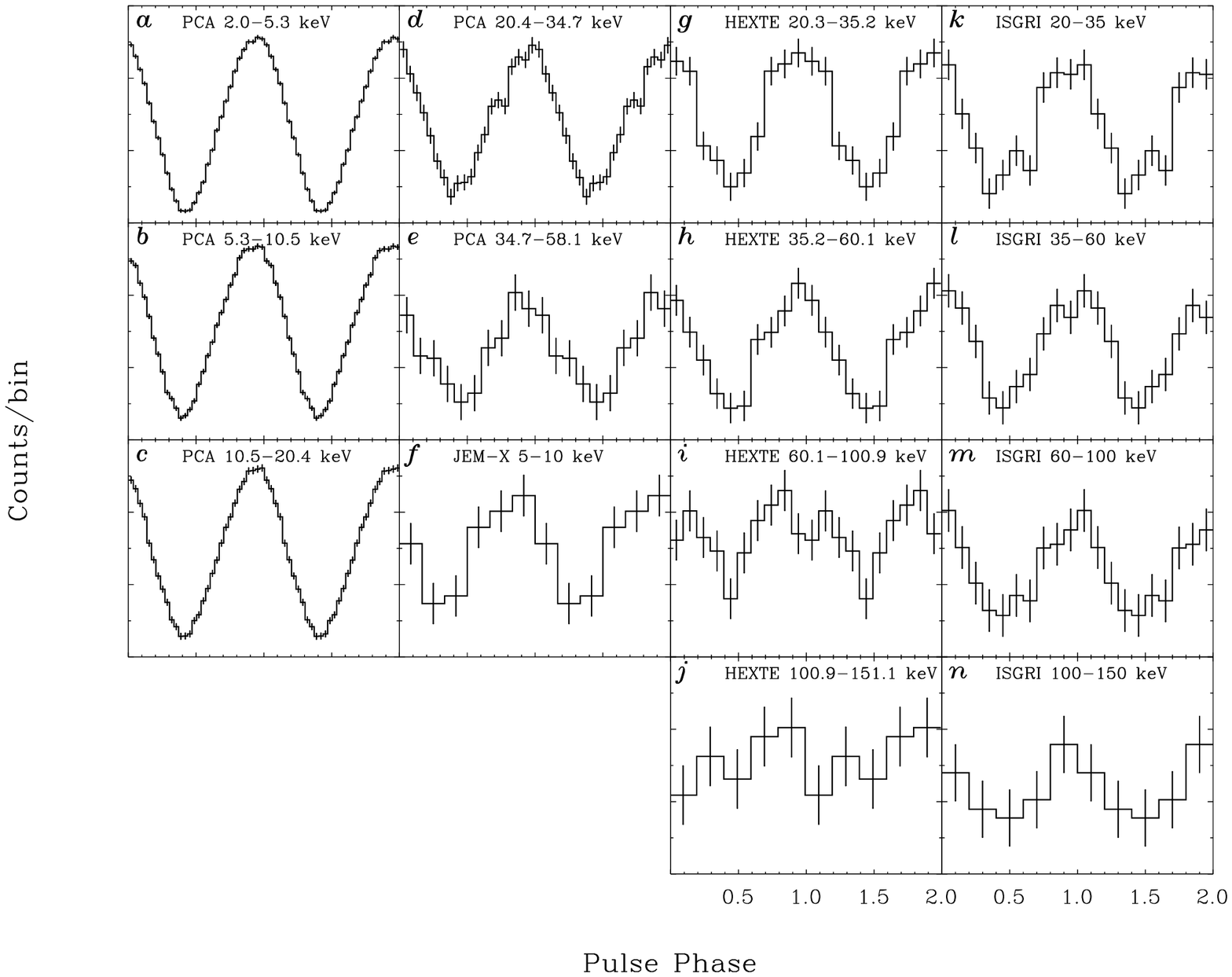}
\vspace{1.2cm}
\caption{  \footnotesize {\it Left:} Simultaneous {\em INTEGRAL}, {\em XMM-Newton}, and {\em RXTE}  spectra of XTE~J1807--294 fitted with an absorbed disk black body, {\sc diskbb}, plus {\sc compps} model (between March 20--22, 2003). The EPIC-pn and MOS2 spectra in the 0.5--10 keV range and PCA/HEXTE in the 3--200 keV and IBIS/ISGRI in 
the 20--200 keV range are shown. The {\sc diskbb} model is shown by a dotted curve, the dashed curve represents the {\sc compps} model, while the total spectrum is shown by a solid line. The lower panel presents the residuals from the best fit. See \cite{mfb05} for more details.
{\it Right:} Pulse-profile collage of IGR~J00291+5934 using data from {\em RXTE}/PCA (2--58 keV; panels a--e), {\em RXTE}/HEXTE (20--151 keV; panels g--j), 
{\em INTEGRAL}/JEM-X (5--10 keV; panel f) and {\em INTEGRAL}/ISGRI (20--150 keV; panels k--n). Two cycles are shown for clarity. All profiles have high 
off-set values due to the intrinsic properties of the instruments. The error bars represent 1 sigma statistical errors. All profiles reach their maximum near phase 
$\sim 0.95$. It is noteworthy the highly sinusoidal shape of the profiles for energies up to 100 keV. }
\label{fig:fig2}
\end{figure} 
 
\section{The AMXP timing above 20 keV}

\begin{figure}[h]
\centering 
\includegraphics[width=0.47\textwidth]{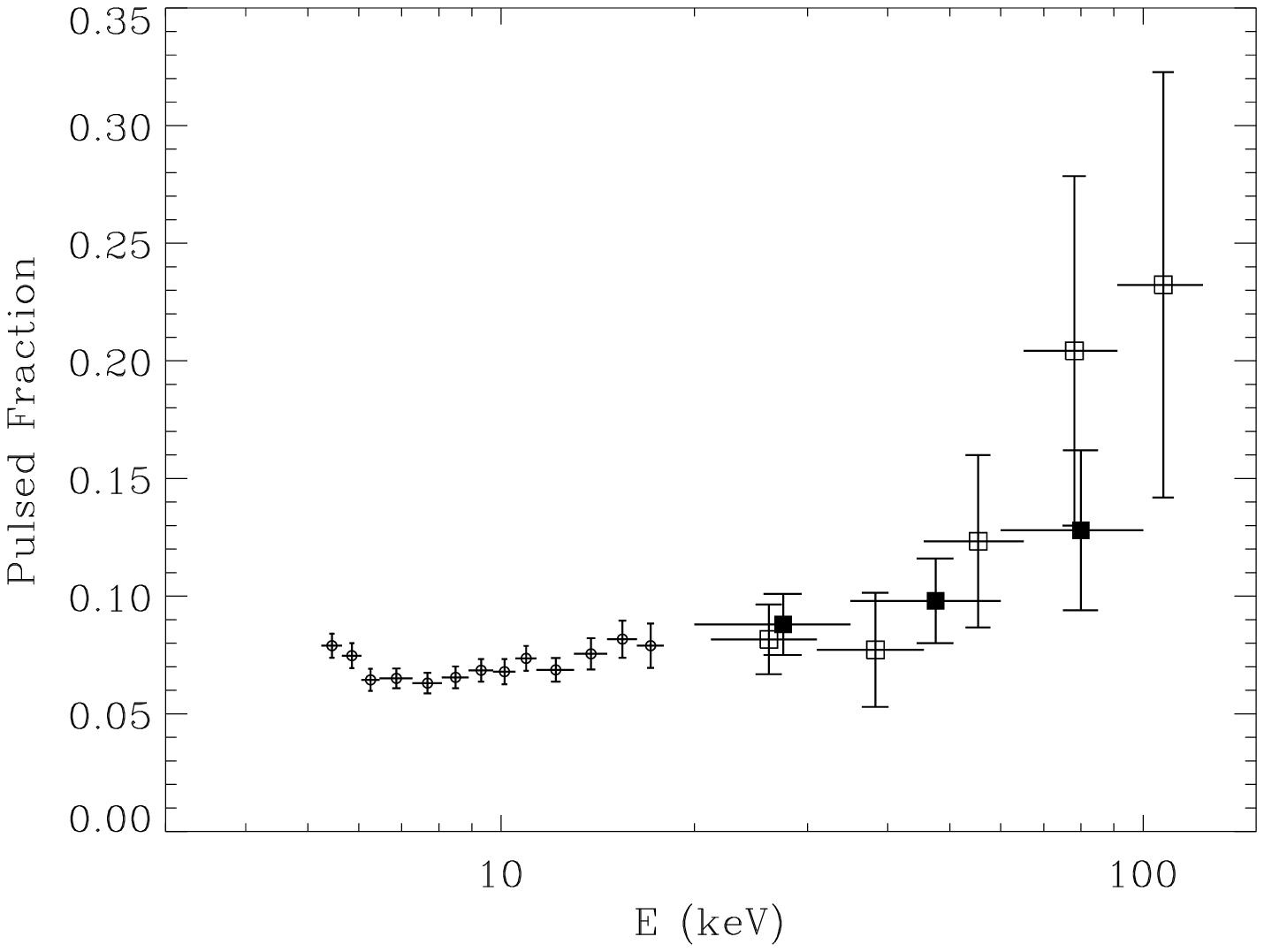}
\hspace{0.55cm}
\includegraphics[width=0.47\textwidth]{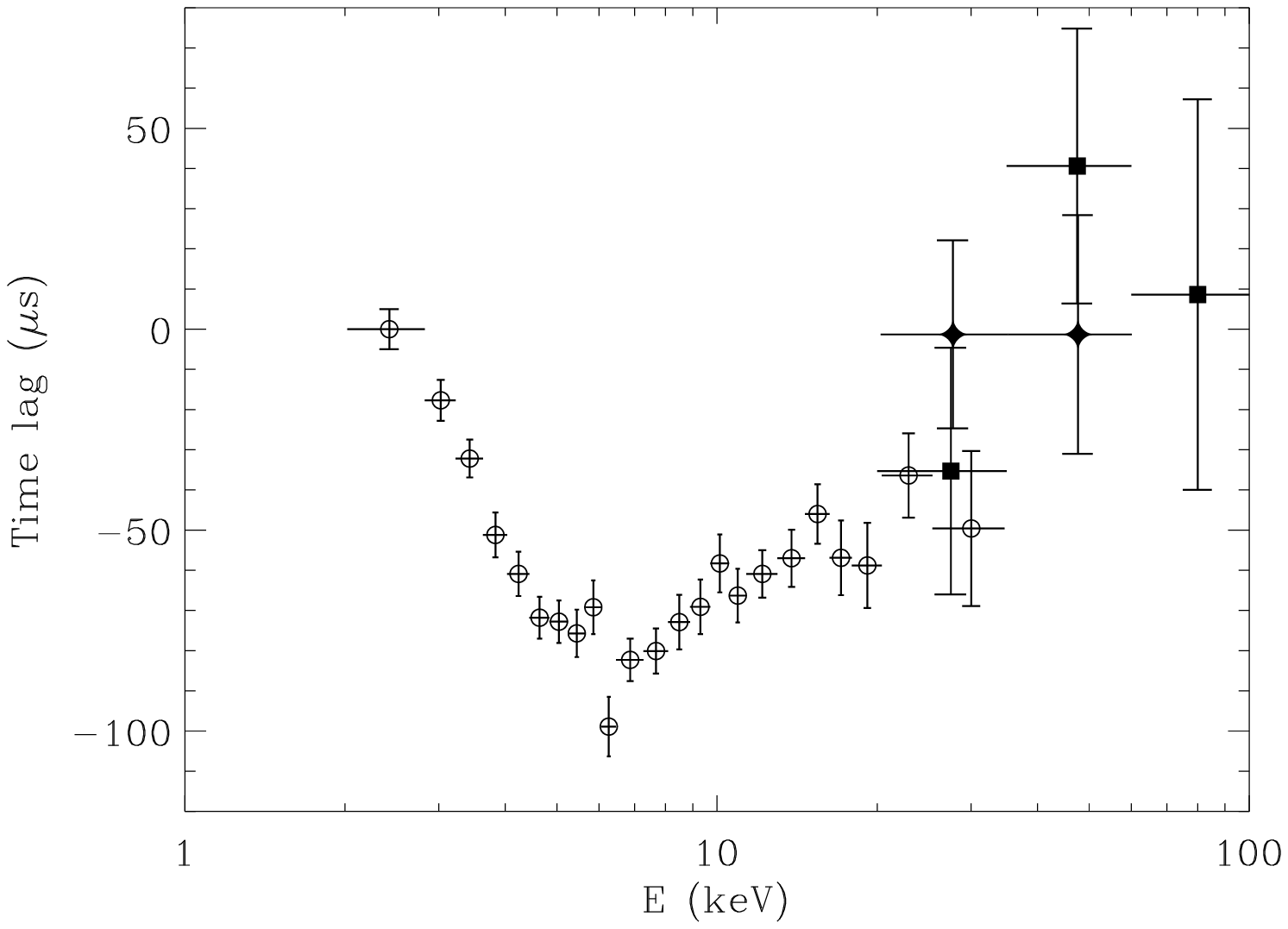}
\caption{  \footnotesize {\it Left panel}:  The pulsed fraction of IGR~J00291+5934 using pulsed/total flux measurements from PCA/HEXTE and  JEM-X/ISGRI (open circles, PCA pulsed flux relative to JEM-X total flux; open squares, HEXTE pulsed flux relative to ISGRI total flux; filled squares, ISGRI pulsed flux relative to ISGRI total flux). The pulsed fraction gradually increases from $\sim 6\%$ at $\sim 6$ keV to $\sim 12-20\%$ at $\sim 100$ keV \cite{mfa05}.
{\it Right panel}: IGR~J00291+5934 time lags as a function of energy in the 2 -- 100 keV energy range combining {\it RXTE}/PCA (2--35 keV, open circles), {\it RXTE}/HEXTE (20-60 keV, filled diamonds), and \inte/ISGRI (20 -- 100 keV, filled squares) measurements.The time lags first increase and then decrease slightly, saturating above 15 keV, and
possibly reaching zero around 50 keV \cite{mfa05}.
}
\label{fig:fig3}
\end{figure} 

In recent years, \inte\ proved to be crucial to perform timing analysis of the high-energy emission from AMXPs in the hard X-ray domain 
($>$20 keV). In the case of IGR~J00291+5934, IGR~J17511--3057, and IGR~J17498--2921, \inte/ISGRI was capable of measuring the pulse 
profile at frequencies of 1.67 ms, 2.5 ms, and 4.08 ms, respectively up to energies of 150\,keV \cite{mfa05,mfalanga11,mfalanga12}. In the case of IGR J00291+5934 see Fig. 2 ({\it right panel}). 
These measurements permitted to perform an analysis of the pulsed fraction up to high X-ray energies and to confirm 
for the first time the increase of the pulsed fraction with energy in an accretion powered AMXP \cite{mfa05}.  The latter could be succesfully interpreted in terms of Doppler effects on the exponentially cut-off Comptonized photons produced from the hot spot on the neutron star surface 
(see Fig. 3; {\it left panel}). Soft phase/time lags (low-energy pulses lag behind the
high-energy pulses) with a complex energy dependence could also be measured and linked to the same effect. According to previous measurements performed on 
the X-ray emission from SAX J1808.4--3658 and XTE~ J1751--350, the time lags revealed by \inte\ monotonically increased
with energy and saturated at about 10--20 keV. The lags are most probably due to the different emission patterns of the blackbody 
and Comptonization components \cite{pg03,gp05,ip09} combined with the action of the 
Doppler effect. This interpretation is supported by the energy dependence of the lags, which 
grows until the contribution of the blackbody becomes negligible. 
At higher energies, the soft lags can evolve further. This is due to the fact that higher-energy 
photons suffer more scatterings and the latter produce notable variations in the emission pattern as a function of the 
energy (see e.g, \cite{mfalanga11,mfalanga12} and reference therein). 
An example of measured time lags in the AMXP IGR~J00291+5934 is shown in Figure 3 ({\it right panel}). 
In this case, the time lags first increase and then decrease slightly, saturating above
15 keV, and possibly reaching zero around 50 keV (see also \cite{FT07}). 

By using {\it RXTE}/PCA data for IGR~J00291+5934 in outburst, it was possible to measure for the first time a 
decrease of the NS spin period while the source was accreting material \cite{mfb05}. This provided an evidence in favor of the so-called ``recycling'' scenario, according to which old neutron stars in LMXB become millisecond X-ray pulsars when they are spun-up to such periods by the transfer of angular momentum from the accreting matter. This result was broadly communicated to scientific community through an ESA press release and global coverage in the popular press (see http://www.esa.int/esaSC/SEMWSAA5QCE\_index\_0.html).

The complementary \inte\ and {\it RXTE} observations provide an opportunity to cross-calibrate the absolute timing of the two instruments. The cross-correlation of the folded solar system barycentered event lists collected during the outburst of IGR~J17511--3057 from ISGRI (15-90 keV, 60 bins), {\it RXTE}/PCA (15-60 keV, 60 bins) and RXTE/HEXTE (15-90 keV, 60 bins) in the same energy band gave time shifts of -49 +/- 114 $\mu$s, and -69 +/- 49 $\mu$s, respectively \cite{Kuiper03,mfa05,mfalanga11,mfalanga12}.
\inte\ event times seemed to be ~ 50 $\mu$s ahead with respect to the {\it RXTE} ones. 
Note, that this is almost the value of 47 $\mu$s that has been added in the process of updating/regenerating the \inte\ time correlation files and used also to update the ISGRI User Manual. So, from an experimental point of view there is no justification to apply this shift of 47 $\mu$s in the \inte\ correlation files, but within errors everything still matches. We can state that \inte\ ISGRI and {\it RXTE} PCA/HEXTE timing measurements are consistent in absolute sense at the 50 $\mu$s level. We have thus the absolute timing information to perform a confident timing analysis. However, only with the {\it RXTE} monitoring observations over long time spans (to establish orbital parameters) an ephemeris could be constructed and used for the \inte/ISGRI events data. A blind searches with \inte\ will not lead to establish the AMXP nature. One long {\it XMM-Newton} observation may establish the AMXP nature, but can not yield an ephemeris without a long monitoring program. Nowadays without monitoring observations and without sensitive X-ray instruments with ms-timing capabilities the discovery of new AMXPs is unlikely.

\section{The AMXP observed X-ray bursts}

\begin{figure}[ht]
\centering 
\includegraphics[angle=-90,width=0.40\textwidth]{fig6.ps}
\hspace{0.55cm}
\includegraphics[width=0.40\textwidth]{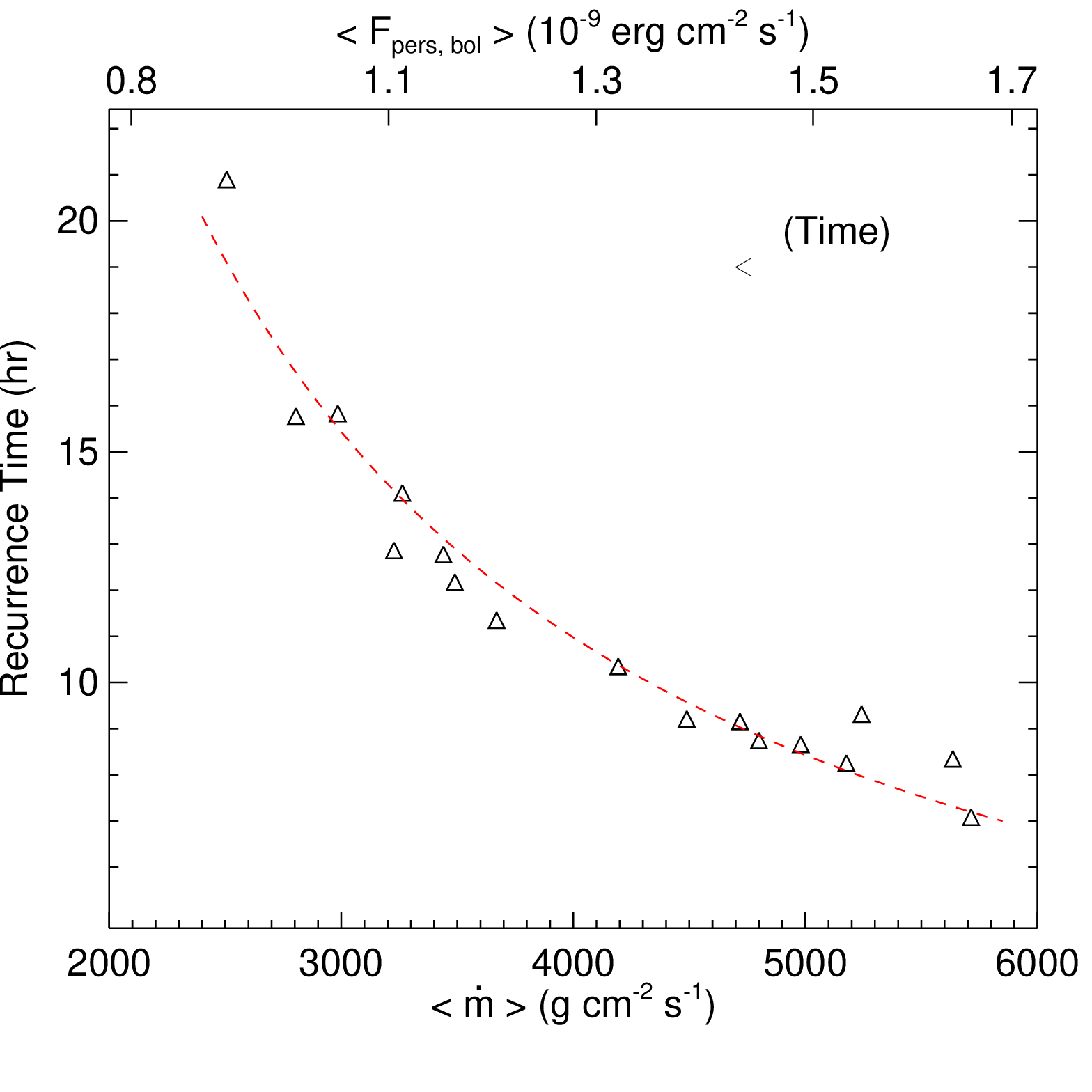}
\caption{  \footnotesize {\it Left panel}: The JEM-X 
  (3--20 keV, upper panel) and IBIS/ISGRI (18--40 keV; lower panel) 
  background subtracted lightcurve of a bright X-ray burst detected from HETE~J1900.1-2455. The time bin is 
  0.5 s. At high-energy the burst shows strong evidence of photospheric radius expansion \cite{mfc07}.
{\it Right panel}: Triangles in the figure represent the observed burst recurrence times shown as a function of local 
mass accretion rate (the corresponding flux is reported in the upper x-axis) for IGR~J17511--3057.  We also show the best-fit 
power-law model. The recurrence time increased with time roughly as $\langle F_{\rm pers,bol}\rangle^{-1.1}$ \cite{mfalanga11}.
}
\label{fig:fig4}
\end{figure} 

\inte\ observed several type-I X-ray bursts from the sources HETE~1900.1--2455, Swift J1749.4--2807, IGR~J17511--3057, and IGR J17498--2921  
\cite{mfc07,ferrigno11,mfalanga11,mfalanga12}. Thermonuclear (type-I)
X-ray bursts are produced by unstable burning of accreted matter on the NS surface. The spectrum from
a few keV to higher energies can usually be well described by using a blackbody model with temperature $kT_{\rm bb}\approx$1--3 keV. 
The energy-dependent decay time of these bursts is attributed to the cooling of the NS photosphere and results in a gradual 
softening of the burst spectrum with time (see e.g., \cite{lewin93} for a review). 
Long ($\sim$220 ks) uninterrupted \inte\ observations of AMXPs in outbursts permitted to detect series of type-I bursts and 
constrain accurately their recurrence time.  In addition, the dependency of the recurrent time on accretion rate and the 
ignition depth could be measured by making use of combined observations carried out by 
\inte, {\it RXTE}, {\it Swift}, {\it Chandra}, and {\it XMM-Newton}. In all the above mentioned sources, the short burst profiles 
provided indication of hydrogen-poor material at ignition, suggesting either that the accreted material is hydrogen-deficient, or that 
the CNO metallicity is slightly higher than the solar value in the AMXPs. For instance, for IGR~J17511--3057 (see Fig. 4 {\it right panel}) 
the variation of the burst recurrence time as a function of $\dot{m}$ (inferred from the X-ray flux) is much smaller than predicted 
by helium-ignition models \cite{mfalanga11}. 

\section{Summary}

\inte\ was able to observe at least one AMXP per year and discovered three objects belonging to this class. By combining high-energy \inte\ data with those obtained from {\it RXTE}, {\it XMM-Newton}, and {\it Swift}, it was possible to study the timing and bursts properties of the AMXPs in a wide energy range (0.3-150 keV).  Our research group has been leading \inte\ observational campaign of AMXPs since the early operational phases. Our AMXP Target of Opportunity (ToO) proposals were approved during AO2-A10 cycles, and during the past few 
years our ToOs were triggered five times. We observed IGR J00291+5934 in 2004, HETE 1900.1--2455 in 2005, IGR J17511--3057 in 2009, IGR J17498--2921 in 2011, and Swift J1749.4--2807 in 2010. These observations were possible thanks to the prompt reactions to our requests from C. Winkler and the {\em INTEGRAL} staff, to whom we are grateful. We are also grateful to all TAC members who positively evaluated our research programs over years and permitted crucial advancements in the understanding of the physics of AMXPs. 

\begin{multicols}{2}
\footnotesize{

}
\end{multicols}
\end{document}